\documentclass[apj]{emulateapj}
\usepackage{tabularx}
\usepackage{multirow}
\usepackage{color}
\definecolor{naviblue}{RGB}{0,0,102}

\begin{document}

\title{DISCOVERY OF X-RAY EMISSION FROM THE FIRST B\lowercase{E}/BLACK HOLE SYSTEM}

\author{P. Munar-Adrover$^{1}$, J. M. Paredes$^{1}$, M. Rib\'o$^{1}$, K. Iwasawa$^{2}$, V. Zabalza$^{3}$, and J. Casares$^{4,5}$}
\affil{$^{1}$Departament d'Astronomia i Meteorologia, Institut de Ci\`encies del Cosmos, Universitat de Barcelona, IEEC-UB, Mart\'{\i} i Franqu\`es 1, E-08028 Barcelona, Spain }
\affil{$^{2}$ICREA, Institut de Ci\`encies del Cosmos, Universitat de Barcelona, IEEC-UB, Mart\'i i Franqu\`es 1, E-08028 Barcelona, Spain}
\affil{$^{3}$Department of Physics and Astronomy, University of Leicester, University Road, Leicester, LE1 7RH, UK}
\affil{$^{4}$Instituto de Astrof\'isica de Canarias, E-38200 La Laguna, Tenerife, Spain}
\affil{$^{5}$Departamento de Astrof\'isica, Universidad de La Laguna, Avda. Astrof\'isico Francisco S\'anchez s/n, E-38271 La Laguna, Tenerife, Spain}

\begin{abstract}

MWC 656 (= HD 215227) was recently discovered to be the first binary system composed of a Be star and a black hole (BH). We observed it with \textit{XMM-Newton}, and detected a faint X-ray source compatible with the position of the optical star, thus proving it to be the first Be/BH X-ray binary. The spectrum analysis requires a model fit with two components, a black body plus a power law, with $k_{\rm B}T = 0.07^{+0.04}_{-0.03}$~keV and a photon index $\Gamma= 1.0\pm0.8$, respectively. The non-thermal component dominates above $\simeq$0.8 keV. The obtained total flux is $F(0.3$--$5.5~{\rm keV}) = (4.6^{+1.3}_{-1.1})\times10^{-14}$ erg cm$^{-2}$ s$^{-1}$. At a distance of $2.6\pm0.6$~kpc the total flux translates into a luminosity $L_{\rm X} = (3.7\pm1.7)\times10^{31}$ erg s$^{-1}$. Considering the estimated range of BH masses to be 3.8--6.9 $M_{\odot}$, this luminosity represents $(6.7\pm4.4)\times10^{-8}~L_{\rm Edd}$, which is typical of stellar-mass BHs in quiescence. We discuss the origin of the two spectral components: the thermal component is associated with the hot wind of the Be star, whereas the power law component is associated with emission from the vicinity of the BH. We also find that the position of MWC~656 in the radio versus X-ray luminosity diagram may be consistent with the radio/X-ray correlation observed in BH low-mass X-ray binaries. This suggests that this correlation might also be valid for BH high-mass X-ray binaries (HMXBs) with X-ray luminosities down to $\sim10^{-8} L_{\rm Edd}$. MWC~656 will allow the accretion processes and the accretion/ejection coupling at very low luminosities for BH~HMXBs to be studied. 

\end{abstract}


\section{Introduction}\label{intro}

MWC~656 was recently discovered to be the first Be star orbited by a black hole (BH) \citep{casares14}. This \textit{solves} the problem of the missing Be/BH binaries \citep{Belczynski07} and opens a window to study the behavior of Be/BH binaries compared to Be/neutron star binaries.

\cite{Lucarelli10} reported the detection of a new unidentified transient point-like source by \textit{AGILE} at energies above 100~MeV, AGL~J2241+4454, with a position uncertainty of 0$^{\circ}$.6 and a flux $F(E > 100~{\rm MeV}) =1.5\times10^{-6}$ ph cm$^{-2}$ s$^{-1}$. However, \textit{Fermi}/LAT did not detect emission at high-energy (HE) in subsequent observations. \cite{Williams10} pointed out the Galactic Be star MWC~656 as the possible optical counterpart. These authors reported an optical photometric periodicity of $60.37\pm0.04$~days for MWC~656, confirmed by \cite{Paredes-Fortuny12}.  \cite{Casares_optical_2012} established the binary nature of the source through optical spectroscopic observations. In a subsequent work, \cite{casares14} presented new data and a double-line solution to the radial velocity curves of the two binary components, allowing them to obtain physical parameters: eccentricity of 0.10$\pm$0.04, periastron at orbital phase $\phi_{\rm per} = 0.01\pm0.10$ (phase 0 is defined at the maximum optical luminosity, corresponding to $T_0 = 2453243.3$ HJD), and mass ratio $M_2/M_1 = 0.41\pm0.07$. They also updated the spectral classification of the Be star to B1.5--B2~III, which implies a Be mass in the range 10--16 $M_{\odot}$ and, therefore, a BH companion of 3.8--6.9 $M_{\odot}$. This makes MWC~656 the first Be binary with a BH companion. The system is located at a distance of $2.6\pm0.6$~kpc \citep{casares14}.

Flux upper limits of MWC 656 have been obtained at different wavelengths. The source was observed in radio using the European VLBI Network (EVN) at 1.6~GHz on 2011 January 25, February 15, and February 28, at orbital phases 0.82, 0.17, and 0.38, respectively (using the ephemeris reported in \citealt{Williams10}). However, the target was not detected during this EVN campaign, with $3\sigma$ flux density upper limits in the range 30--66 $\mu$Jy \citep{Moldon_thesis}.

{In the X-ray domain, MWC~656 was observed by \textit{ROSAT} in 1993 July 7--11 at 0.1--2.4~keV energies for a short time ($\sim$3--6~ks each day). \cite{casares14} obtained a 90\% confidence level (c.l. hereafter) flux upper limit of $F(0.1$--$2.4~{\rm keV})<1.2\times 10^{-13}$ erg~cm$^{-2}$~s$^{-1}$, assuming a photon index $\Gamma=2.0$ (typical of BH X-ray binaries (XRBs) in the quiescent state; \citealt{Plotkin13}) and an interstellar hydrogen column density $N_{\rm H}= 1.4\times 10^{21}~\rm{cm}^{-2}$. In addition, two 1~ks \textit{Swift}/X-Ray Telescope (XRT) observations performed on 2011 March 8 provide a 90\% c.l.\ flux upper limit in the 0.3--10~keV energy range of $F(0.3$--$10.0~{\rm keV})<4.6\times 10^{-13}$ erg~cm$^{-2}$~s$^{-1}$ \citep{casares14}. The \textit{ROSAT} and \textit{Swift}/XRT observations were carried out at orbital phases 0.63--0.70 and 0.52, respectively.

{In the HE gamma-ray domain we took all the available \textit{Fermi}/LAT data and analyzed it, finding no evidence of a detection and deriving a 95\% c.l. upper limit of $F(E > 100~{\rm MeV}) < 7.9\times10^{-10}$ cm$^{-2}$ s$^{-1}$. Finally, MWC~656 has been observed at very high energies using the MAGIC Telescopes in 2012 May and June, corresponding to orbital phase intervals 0.83--0.95 and 0.20--0.28, respectively. The source was not detected in any of these observations. Preliminary integral upper limits were set at 95$\%$ c.l., assuming a spectral index $\Gamma=2.5$ at a level of $F(E>300 {\rm ~GeV}) < 2.1\times$10$^{-12}$cm$^{-2}$s$^{-1}$ \citep{MAGIC_ICRC2013}.

In the present work we show the results obtained from our recent \textit{XMM-Newton} observations. We detect a faint source coincident with the position of MWC~656, making it the first Be/BH binary ever detected in X-rays and thus confirming it as a high-mass X-ray binary (HMXB). The spectrum can be fitted with a thermal plus a non-thermal component. We discuss our results in the context of wind emission from massive stars, the quiescent state of BH XRBs, and the empirical BH radio/X-ray correlation. This Letter is organized as follows: we present the \textit{XMM-Newton} observations and analysis in Section~\ref{sect_obs}, results obtained and their discussion in Sections~\ref{sect_results} and ~\ref{sect_discussion}, and finish by presenting our conclusions in Section~\ref{sect_conclusions}.


\section{X-ray observations and analysis}\label{sect_obs}

\begin{figure}[t!]
\resizebox{\hsize}{!}{\includegraphics{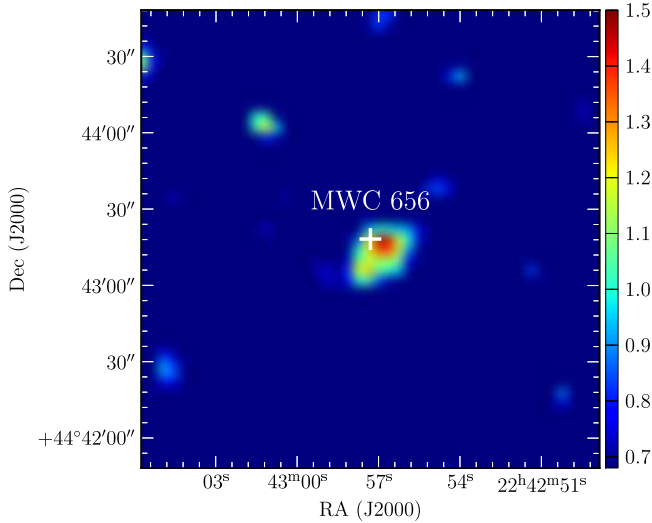}}  
  \caption{EPIC-pn camera image at the position of MWC~656 in the 0.3--5.5 keV energy band smoothed using a Gaussian interpolation with a $2^{\prime \prime}$ kernel. \label{Ximage}}
\end{figure}

We present here new X-ray observations of MWC~656 performed with \textit{XMM-Newton} for 14~ks\footnote{The observation ID is 0723610201}. The observations were carried out in pointing mode in 2013 June 4, when the system was at orbital phase 0.08 (close to periastron). The medium thickness optical blocking filter and Full Frame mode were used in the three EPIC detectors (pn, MOS1 and MOS2) for the imaging observation. Analysis of the data was done using the \textit{XMM-Newton} Science Analysis System (SAS) version 12.0.1 and the set of {\tt ftools} from HEASOFT version 6.14\footnote{http://heasarc.gsfc.nasa.gov/}. In a first step we cleaned the event files of the three EPIC detectors, removing the flaring high background periods. For this purpose we selected the Good Time Intervals in which the count rate for the most energetic events ($E\geq$~10~keV) was below the standard threshold for each detector. After this cleaning process, the observation time that remained in each of the three detectors was $10.1$ ks, $13.4$ ks, and $13.5$ ks for pn, MOS1, and MOS2, respectively.

 A search for sources was performed using the {\tt edetect-chain} command in SAS, which concatenates a series of tasks that produces exposure maps, detector mask images, background maps, detected source lists, and sensitivity maps. As a result, a list of detected sources, including MWC~656, was obtained, which contained count-rates, fluxes and positions, for each detected source, among other information. We used the cleaned event files for spectral analysis. The source spectrum was extracted from a $30^{\prime\prime}$ radius circle centered on the position of MWC~656 whereas the background spectrum was extracted from a source-free region on the same CCD chip, using a circle of $90^{\prime\prime}$ radius. Response and ancillary files were obtained with the SAS tools {\tt rmfgen} and {\tt arfgen}, respectively.


\section{X-ray results}\label{sect_results}

\begin{figure}[t!]
\resizebox{\hsize}{!}{\includegraphics{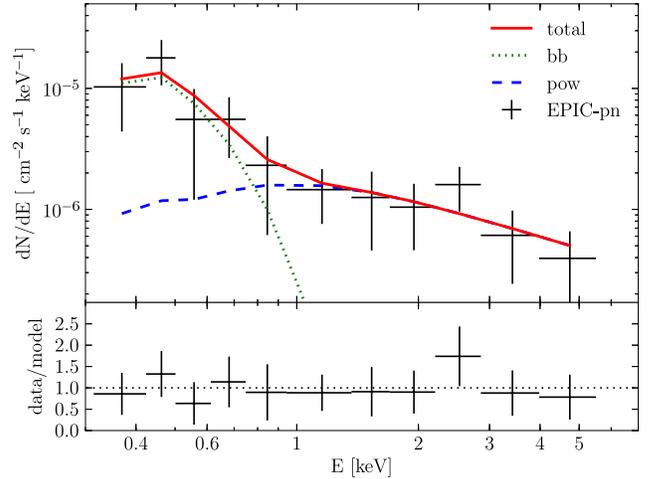}}
\caption{MWC~656 \textit{XMM-Newton} EPIC-pn spectrum in the 0.3--5.5 keV energy range (data points) overplotted with the fitted absorbed black-body (green dotted line) plus a power law (blue dashed line) model. The red solid line represents the total flux. The lower panel illustrates the ratio between the observational data and the total model.\label{hd_spec}}
\end{figure}
\begin{table*}[t!]

  \begin{center}
    \caption{X-Ray Spectral Fit Parameters of MWC~656 \label{tab_spec_fit}}        
    \scalebox{1}[1]{
    \begin{tabular}{ccccccccl}
    \hline
    \hline
\multirow{2}{*}{Model$^a$} & \multicolumn{2}{c}{Parameters} & & \multicolumn{3}{c}{$F(0.3$--$5.5~{\rm keV})$/$10^{-14}$ (erg cm$^{-2}$ s$^{-1}$)} & \multirow{2}{*}{C-statistic} &\multirow{2}{*}{Comments}\\
\cline{2-3}
\cline{5-7}
           & $k_{\rm B}T$ [keV] & $\Gamma$    & & thermal  & non-thermal & total & &     \\
\hline
pow        &         $-$        & $2.0^{+1.0}_{-0.8}$ & & $-$           & $2.3^{+0.8}_{-0.7}$ & $2.3^{+0.8}_{-0.7}$  &14.6 & Deviations at $\sim$0.5, $\sim$1.0, $\sim$2.5 keV \\
bb         &    $0.12^{+0.07}_{-0.05}$   & $-$         & & $2.0^{+0.9}_{-0.8}$ & $-$         & $2.0^{+0.9}_{-0.8}$ & 31.4 & No good fit above 1.5 keV \\
bb+pow     &    $0.07^{+0.04}_{-0.03}$   & $1.0\pm0.8$ & & $2.6^{+3.0}_{-1.4}$ & $2.0^{+0.8}_{-0.7}$ & $ 4.6^{+1.3}_{-1.1}$  & 2.8 & Good fit (used in this work) \\
diskbb+pow &    $0.09^{+0.04}_{-0.06}$   & $1.0\pm0.8$ & & $2.7^{+3.4}_{-1.4}$ & $2.0^{+0.8}_{-0.7}$ & $ 4.7^{+1.4}_{-1.1}$  & 2.8 & Good fit \\
\hline
\multicolumn{9}{l}{Note. $a$: Models correspond to powerlaw (pow), black body (bb) and multi-temperature disk black body (diskbb)} \\
     \end{tabular}}
  \end{center}
\end{table*} 


We detect a faint source at $4.4\sigma$ c.l. coincident with MWC~656 (see Figure \ref{Ximage}). The X-ray position is R.A.$=22^{\rm h}42^{\rm m}57^{\rm s}.1$, decl.$=44^{\circ}43^{\prime}13^{\prime\prime}$, with a $3\sigma$ uncertainty radius of $7^{\prime\prime}$. This is compatible with the \textit{Hipparcos} position of MWC~656 at $2.4\sigma$. The source is only detected in the low energy range of the EPIC-pn detector, between 0.3 and 5.5 keV. In the MOS1 and MOS2 instruments MWC~656 appears as a faint excess below $3\sigma$ c.l., probably due to the lower effective area at these energies. 

The EPIC-pn spectrum (see Figure \ref{hd_spec}) was binned so that there were at least 10 counts per bin. The binned spectrum was analyzed using XSPEC version 12.8.1 \citep{Arnaud96}. The low number of counts required the use of the {\tt cstat} statistic within XSPEC to estimate the best-fit parameters and their associated uncertainties \citep{Cash1979}. In all fits we introduced a fixed interstellar photoelectric absorption through the {\tt wabs} model within XSPEC, with $N_{\rm H} = 1.8\times10^{21}$ cm$^{-2}$. This value was computed using the relations $N_{\rm H} = 2.7\times10^{21} A_{\rm V}$ \citep{Nowak2012}, and $N_{\rm H} = (2.21\pm0.09)\times10^{21} A_{\rm V}$ \citep{Guever2009} and taking the average value. For the extinction we used $A_{\rm V}=0.74\pm0.19$ mag computed using $E(B-V)=0.24\pm0.06$ \citep{casares14} and assuming $R=A_{\rm V}/E(B-V)=3.1$.

A power law (pow) model fit yields a photon index $\Gamma = 2.0^{+1.0}_{-0.8}$, similar to that found in other BH XRBs. However, the model presents strong deviations from the data at $\sim$0.5, $\sim$1.0, and $\sim$2.5 keV. A black body (bb) model is also unable to reproduce the observed spectrum above 1.5~keV. For this reason we decided to fit the data with a model including two components: a black-body plus a power law. This fit yields $k_{\rm B}T=0.07^{+0.04}_{-0.03}$~keV, photon index $\Gamma = 1.0\pm0.8$, and a total unabsorbed flux $F(0.3$--$5.5~{\rm keV}) = (4.6^{+1.3}_{-1.1})\times10^{-14}$ erg cm$^{-2}$ s$^{-1}$. The contribution of each of the two components to the total flux is $F_{\rm bb}=(2.6^{+3.0}_{-1.4})\times10^{-14}$ erg cm$^{-2}$ s$^{-1}$ and $F_{\rm pow}=(2.0^{+0.8}_{-0.7})\times10^{-14}$ erg cm$^{-2}$ s$^{-1}$. A solution swapping the two components, with the power law at low energies and the black body at higher energies, is also possible, although in this case the photon index is unconstrained. On the other hand, a fit using a multi-temperature disk black body (diskbb, \citealt{Mitsuda1984}) plus a power law gives similar results. 

We summarize the results of the most relevant fits in Table \ref{tab_spec_fit}, where the fluxes are unabsorbed. The uncertainties are quoted at the 1$\sigma$ level. Lower values of the C-statistic reflect better fits \citep{Cash1979}. The total fluxes obtained in the 0.3--5.5~keV energy range in all fits are around one order of magnitude lower than the upper limits obtained from previous \textit{ROSAT} and \textit{Swift}/XRT observations \citep{casares14}. It is clear from the spectral analysis that two components are required to fit our data (see Figure \ref{hd_spec}). The fit with a black body plus a power law is more realistic and simple and we will consider it throughout the next sections. We note that a joint fit using both the EPIC-pn and the EPIC-MOS1 and EPIC-MOS2 data sets provided the same results within uncertainties.

We searched for a possible extension of the 0.3--5.5~keV point source by adjusting radial profiles to the EPIC-pn image. However, no significant extension was found. We also searched for variability during the observation by obtaining a background subtracted light curve of the source region but no significant variability was found.


\section{Discussion}\label{sect_discussion}

Our detection of the X-ray counterpart of MWC~656 allows us to classify the Be/BH system as an HMXB, the first XRB of this type. 

The spectrum is best fit with a model that includes a thermal and a non-thermal component, with the non-thermal component dominating above $\simeq$0.8 keV. Given the source distance of 2.6$\pm$0.6~kpc, the total X-ray luminosity is $L_{\rm X}=(3.7\pm1.7)\times 10^{31}$ erg s$^{-1} $ in the 0.3--5.5 keV band. This luminosity represents $(6.7\pm4.4)\times 10^{-8} L_{\rm Edd}$ for the estimated range of BH masses of 3.8--6.9 M$_{\odot}$. The thermal and non-thermal contributions are $L_{\rm bb} = (2.1^{+2.8}_{-1.5})\times 10^{31}$ erg s$^{-1}$ and $L_{\rm pow}=(1.6^{+1.0}_{-0.9})\times 10^{31}$ erg s$^{-1}$, respectively.

\subsection{Origin of the Thermal Component}

The X-ray spectrum in B-type stars is typically represented by either a hot thermal component at about 1 keV or the sum of two thermal components, one around 0.4 keV and another one around 2 keV \citep{Naze09}. In our case the best fit yields a temperature of $k_{\rm B}T = 0.07^{+0.04}_{-0.03}$ keV, which is not far from that observed in B-type stars. We can also compare the results of our fit with the correlation $L_{\rm X} \sim 10^{-7}L_{\rm bol}$ \citep[see][]{Berghoefer97,Cohen97}, where $L_{\rm X}$ represents the thermal X-ray luminosity. The bolometric luminosity of MWC~656 is $7\times 10^{37}$ erg s$^{-1}$, considering $M_{\rm V} = -4.1$ and a bolometric correction of $-$1.8 according to \cite{Straizys81}. In our case, given the derived thermal X-ray luminosity, we find $L_{\rm X}/L_{\rm bol}=3\times 10^{-7}$, compatible within uncertainties with the correlation quoted above. Although BH XRBs show a thermal component at high accretion rates, this component disappears in quiescence both due to the low accretion rate and to the radiatively inefficient nature of the accretion disk, and only a soft power law with $\Gamma\sim2$ is detected. Therefore, the detected thermal component is not expected to arise in the vicinity of the BH. Hence, the obtained results suggest that the thermal component of our X-ray spectrum arises from the hot wind of the Be star.

\begin{figure*}[t!]
\begin{center}
\resizebox{0.9\hsize}{!}{\includegraphics{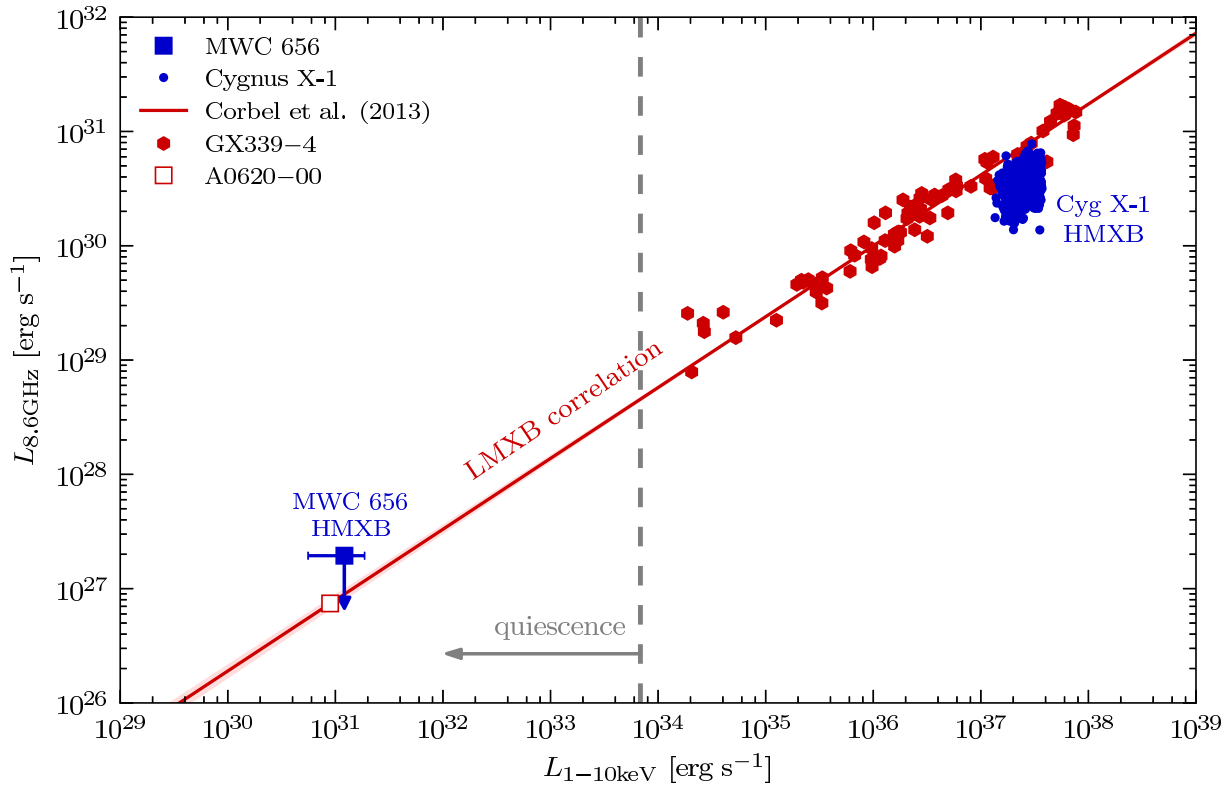}}
\caption{Radio vs X-ray luminosity diagram including the position of MWC 656 (blue square) according to non-simultaneous X-ray observation (this work; accounting only for the non-thermal contribution) and the lowest radio flux density upper limit from \cite{Moldon_thesis}. The small blue dots indicate the region of the parameter space where Cygnus X-1 has been detected in the low/hard state \citep{Gallo12}. We also plot the radio/X-ray correlation for BH LMXBs of \cite{Corbel13} (red solid line plus light red shadow), together with data on the BH LMXBs GX 339$-$4 (red hexagons) and A0620$-$00 (empty red square) to display the luminosity range of real sources. The gray dashed line separates the quiescent state region (left) from the other states (right) according to the threshold set by \cite{Plotkin13}. The position of MWC 656 is very close to the one of the LMXB A0620$-$00 in quiescence, indicating that the radio/X-ray correlation might also be valid for BH HMXBs down to very low luminosities.\label{correlation}}
\end{center}
\end{figure*}

\subsection{Origin of the Non-thermal Component}

BH low-mass X-ray binaries (LMXBs) display a positive correlation between the radio and X-ray luminosities which is thought to be due to an accretion disk--jet coupling. This coupling is invoked to explain the different state transitions observed in BH binary systems \citep{Fender10}. At the beginning of the outburst and during the outburst decay, the source is in a hard state \citep{RemillardMcClintock06}. This hard state is characterized by a power law spectrum with photon index $\Gamma\sim1.5$, a luminosity below 0.1 $L_{\rm Edd}$, and the presence of an active jet in the radio domain with an almost flat radio spectrum. The X-ray emission is non-thermal and thought to be due to Comptonization in a corona near the BH \citep{Zdziarski98,Nowak99} or to optically thin synchrotron emission from the base of the jet \citep{Markoff05}. At the end of the outburst the X-ray emission decays again to a low/hard state below 0.01--0.04 $L_{\rm Edd}$ and eventually to a quiescent state below $10^{-5}~L_{\rm Edd}$ \citep{Plotkin13}. The quiescent X-ray spectrum is reproduced by a power law with an average photon index $\Gamma=2.08\pm0.07$ \citep{Plotkin13}. As in the low--hard state, the quiescent spectrum is non-thermal emission from the corona or the jet base, but slightly softer. 

The non-thermal X-ray luminosity of MWC~656 in the 0.3--5.5~keV band is $L_{\rm X}=(3.1\pm2.3)\times 10^{-8} L_{\rm Edd}$ for the estimated BH mass of 3.8--6.9 $M_{\odot}$ \citep{casares14}. This luminosity is around three orders of magnitude below the $10^{-5}~L_{\rm Edd}$ threshold from \cite{Plotkin13}, making our results compatible with MWC~656 being in quiescence.

In addition, the best fit model gives a photon index $\Gamma = 1.0\pm0.8$ which is roughly compatible with the results of \cite{Plotkin13} for BH LMXBs in quiescence. Nevertheless, poor statistics in the spectrum prevent us to constrain the photon index of MWC 656. 

\subsection{MWC~656 Within the BH Radio/X-ray Correlation}

We can compare the X-ray flux obtained in this Letter and the previous radio flux density upper limits with the radio/X-ray correlation from \cite{Corbel13} (see Figure \ref{correlation}). The EVN radio flux density $3\sigma$ upper limits are in the range 30--66 $\mu$Jy \citep{Moldon_thesis}, although they are not simultaneous nor at the same orbital phase as our \textit{XMM-Newton} observation. Converting the lowest radio flux density upper limit into luminosity we obtain $L_{\rm r} < 2\times10^{27}$ erg s$^{-1}$ at 8.6~GHz assuming a flat radio spectrum. The X-ray luminosity corresponding to the contribution of the power law component of our fit, once extrapolated to the 1.0--10 keV energy band, is $L_{\rm X}=(1.2\pm0.6)\times 10^{31}$ erg s$^{-1}$. MWC 656 is located in the lower-left side of the luminosity diagram (see Figure~\ref{correlation}), very close to A0620$-$00 and just above the correlation from \cite{Corbel13}. The expected radio luminosity considering our X-ray flux measurement and the \cite{Corbel13} correlation is $L_{\rm r}=(8.8\pm3.1)\times 10^{26}$ erg s$^{-1}$, which translates into a radio flux density of $S_{\nu} = 13_{-4}^{+5}~\mu{\rm Jy}$ at 8.6~GHz (or $8_{-4}^{+6}$ $\mu$Jy using the correlation from \citealt{Gallo12}). The quoted uncertainties account for the X-ray flux, distance, and correlation parameters uncertainties. This result is not far from the upper limits we have obtained up to now.

\cite{Gallo07} found a significant number of outliers in the original radio/X-ray best-fitting relation \citep{Gallo03}, indicating that there might be a second track in the high X-ray luminosity region with a steeper slope. \cite{Coriat11} suggested that these two tracks may collapse into a single track for X-ray luminosities below $3\times10^{-4} L_{\rm Edd}$, as it seems to occur in the case of H1743--322. \cite{Gallo12} studied the case in detail adding newly discovered sources, as well as new data from well-studied ones, and concluded that the two--track scenario better explains the observational data. The radio flux density upper limits for MWC~656 are consistent with both tracks (assuming the lower one also extends toward lower X-ray luminosities). For the lower track the expected radio luminosity would be undetectable by the current radio telescopes.

So far, the only known HMXB containing a confirmed BH in our Galaxy is Cygnus X-1. This system has an X-ray luminosity in the range $(1.0$--$7.0)\times10^{37}$ erg s$^{-1}\equiv(0.8$--$3.0)\times10^{-2}$ $L_{\rm Edd}$ for a $\sim 15$ $M_{\odot}$ BH in the 1--10~keV range (\citealt{Gallo12}; but see \citealt{Zdziarski12b}). Its radio and X-ray fluxes also fulfill the radio/X-ray correlation during the hard state, similar to what is observed in LMXBs \citep{Zdziarski12b}. In contrast, the low X-ray luminosity of MWC~656 makes it comparable to the faintest LMXBs, such as A0620--00 (see Figure \ref{correlation} and \citealt{Plotkin13}), allowing the study accretion processes and the accretion/ejection coupling at very low luminosities for BH HMXBs to be also studied.


\section{Conclusions}\label{sect_conclusions}

We have detected the X-ray counterpart of the first BH orbiting a Be star, MWC~656, confirming it to be an XRB and thus classifying it as an HMXB. Due to the low number of counts we cannot fully characterize its spectrum, although thermal and non-thermal components seem to be required to explain the low-energy and high-energy part of the X-ray spectrum, respectively. These two components are interpreted as the contribution from the hot wind of the Be star and the emission close to the BH, respectively. The non-thermal X-ray flux translates into a luminosity well below the threshold set by \cite{Plotkin13} for quiescent BH binaries. Using the EVN radio flux density upper limits and our X-ray luminosity we find that MWC 656 is located in the lower-left side of the luminosity diagram, in a region where it may be consistent with and just above the correlation from \cite{Corbel13}. Consequently, the radio/X-ray correlation might also be valid for BH HMXBs. In this context, MWC 656 will allow the study of accretion processes and of accretion/ejection coupling at very low luminosities for BH HMXBs. Further deep X-ray observations are needed to better characterize the spectrum and constrain the spectral parameters to allow better interpretation. Deep simultaneous radio observations are needed to study the low luminosity accretion/ejection coupling in BH HMXBs.


\acknowledgements
We thank the anonymous referee for very useful comments. We also thank S.\ Corbel for providing the data for LMXBs. We acknowledge support by DGI of the Spanish Ministerio de Econom\'{\i}a y Competitividad (MINECO) under grants AYA2010-21782-C03-01, AYA2010-18080 and FPA2010-22056-C06-02. P.M-A acknowledges financial support from the Universitat de Barcelona through an APIF fellowship. J.M.P. acknowledges financial support from ICREA Academia.


\begin{thebibliography}{19}
\expandafter\ifx\csname natexlab\endcsname\relax\def\natexlab#1{#1}\fi

\bibitem[{{Arnaud}(1996)}]{Arnaud96}
{Arnaud}, K.~A. 1996, in ASP Conf. Ser.,
  101, Astronomical Data Analysis Software and Systems V, 
  ed. G.~H. {Jacoby} \& J.~{Barnes} (San Francisco, CA: ASP), 17

\bibitem[{{Belczynski} \& {Ziolkowski}(2009)}]{Belczynski07}
{Belczynski}, K. \& {Ziolkowski}, J. 2009, \apj, 707, 870

\bibitem[{{Berghoefer} {et~al.}(1997){Berghoefer}, {Schmitt}, {Danner}, \&
  {Cassinelli}}]{Berghoefer97}
{Berghoefer}, T.~W., {Schmitt}, J.~H.~M.~M., {Danner}, R., \& {Cassinelli},
  J.~P. 1997, \aap, 322, 167

\bibitem[{{Casares} {et~al.}(2014){Casares}, {Negueruela}, {Rib{\'o}}, {Ribas},
  {Paredes}, {Herrero}, \& {Sim{\'o}n-D{\'{\i}}az}}]{casares14}
{Casares}, J., {Negueruela}, I., {Rib{\'o}}, M., {et~al.} 2014, Nat, 505, 378

\bibitem[{{Casares} {et~al.}(2012){Casares}, {Rib{\'o}}, {Ribas}, {Paredes},
  {Vilardell}, \& {Negueruela}}]{Casares_optical_2012}
{Casares}, J., {Rib{\'o}}, M., {Ribas}, I., {et~al.} 2012, \mnras, 421, 1103

\bibitem[{{Cash}(1979)}]{Cash1979}
{Cash}, W. 1979, \apj, 228, 939

\bibitem[{{Cohen} {et~al.}(1997){Cohen}, {Cassinelli}, \&
  {Macfarlane}}]{Cohen97}
{Cohen}, D.~H., {Cassinelli}, J.~P., \& {Macfarlane}, J.~J. 1997, \apj, 487,
  867

\bibitem[{{Corbel} {et~al.}(2013){Corbel}, {Coriat}, {Brocksopp}, {Tzioumis},
  {Fender}, {Tomsick}, {Buxton}, \& {Bailyn}}]{Corbel13}
{Corbel}, S., {Coriat}, M., {Brocksopp}, C., {et~al.} 2013, \mnras, 428, 2500

\bibitem[{{Coriat} {et~al.}(2011){Coriat}, {Corbel}, {Prat}, {Miller-Jones},
  {Cseh}, {Tzioumis}, {Brocksopp}, {Rodriguez}, {Fender}, \&
  {Sivakoff}}]{Coriat11}
{Coriat}, M., {Corbel}, S., {Prat}, L., {et~al.} 2011, \mnras, 414, 677

\bibitem[{{Fender}(2010)}]{Fender10}
{Fender}, R. 2010, Lecture Notes in Physics, Vol. 794, Berlin Springer Verlag, ed. T.~{Belloni}, 115

\bibitem[{{Gallo}(2007)}]{Gallo07}
{Gallo}, E. 2007, AIP Conf. Proc. 924, The Multicolored 
 Landscape of Compact Objects and Their Explosive Origins
  ed. T.~{di Salvo}, G.~L. {Israel}, L.~{Piersant}, L.~{Burderi}, G.~{Matt},
  A.~{Tornambe}, \& M.~T. {Menna} (Melville, NY: AIP), 715

\bibitem[{{Gallo} {et~al.}(2003){Gallo}, {Fender}, \& {Pooley}}]{Gallo03}
{Gallo}, E., {Fender}, R.~P., \& {Pooley}, G.~G. 2003, \mnras, 344, 60

\bibitem[{{Gallo} {et~al.}(2012){Gallo}, {Miller}, \& {Fender}}]{Gallo12}
{Gallo}, E., {Miller}, B.~P., \& {Fender}, R. 2012, \mnras, 423, 590

\bibitem[{{G{\"u}ver} \& {{\"O}zel}(2009)}]{Guever2009}
{G{\"u}ver}, T. \& {{\"O}zel}, F. 2009, \mnras, 400, 2050

\bibitem[{{L{\'o}pez-Oramas} {et~al.}(2013){L{\'o}pez-Oramas}, {Blanch Bigas},
  {Cortina}, {Hadasch}, {Herrero}, {Marcote}, {Munar-Adrover}, {Mold{\'o}n},
  {Paredes}, {Ribas}, {Rib{\'o}}, {Torres}, {R.~Zanin for the MAGIC
  COLLABORATION}, {Casares}, \& {Rea}}]{MAGIC_ICRC2013}
{L{\'o}pez-Oramas}, A., {Blanch Bigas}, O., {Cortina}, J., {et~al.} 2013, in International Cosmic Ray Conference 2013 Proceedings, Observations of VHE Gamma-ray binaries with the MAGIC Telescopes (arXiv:1311.5711)

\bibitem[{{Lucarelli} {et~al.}(2010){Lucarelli}, {Verrecchia}, {Striani},
  {Pittori}, {Tavani}, {Vercellone}, {Bulgarelli}, {Gianotti}, {Trifoglio},
  {Chen}, {Giuliani}, {Mereghetti}, {Caraveo}, {Perotti}, {Donnarumma},
  {D'Ammando}, {Del Monte}, {Evangelista}, {Feroci}, {Lazzarotto}, {Pacciani},
  {Soffitta}, {Costa}, {Lapshov}, {Rapisarda}, {Argan}, {Piano}, {Pucella},
  {Sabatini}, {Trois}, {Vittorini}, {Fuschino}, {Galli}, {Labanti},
  {Marisaldi}, {Di Cocco}, {Pellizzoni}, {Pilia}, {Barbiellini}, {Longo},
  {Moretti}, {Vallazza}, {Morselli}, {Picozza}, {Prest}, {Lipari}, {Zanello},
  {Cattaneo}, {Rappoldi}, {Santolamazza}, {Colafrancesco}, {Giommi}, \&
  {Salotti}}]{Lucarelli10}
{Lucarelli}, F., {Verrecchia}, F., {Striani}, E., {et~al.} 2010, ATel, 2761, 1

\bibitem[{{Markoff} {et~al.}(2005){Markoff}, {Nowak}, \& {Wilms}}]{Markoff05}
{Markoff}, S., {Nowak}, M.~A., \& {Wilms}, J. 2005, \apj, 635, 1203

\bibitem[{{Mitsuda} {et~al.}(1984){Mitsuda}, {Inoue}, {Koyama}, {Makishima},
  {Matsuoka}, {Ogawara}, {Suzuki}, {Tanaka}, {Shibazaki}, \&
  {Hirano}}]{Mitsuda1984}
{Mitsuda}, K., {Inoue}, H., {Koyama}, K., {et~al.} 1984, \pasj, 36, 741

\bibitem[{{Mold\'on}(2012)}]{Moldon_thesis}
{Mold\'on}, J. 2012, PhD thesis, Univ. de Barcelona

\bibitem[{{Naz{\'e}}(2009)}]{Naze09}
{Naz{\'e}}, Y. 2009, \aap, 506, 1055

\bibitem[{{Nowak} {et~al.}(2012){Nowak}, {Neilsen}, {Markoff}, {Baganoff},
  {Porquet}, {Grosso}, {Levin}, {Houck}, {Eckart}, {Falcke}, {Ji}, {Miller}, \&
  {Wang}}]{Nowak2012}
{Nowak}, M.~A., {Neilsen}, J., {Markoff}, S.~B., {et~al.} 2012, \apj, 759, 95

\bibitem[{{Nowak} \& {Wilms}(1999)}]{Nowak99}
{Nowak}, M.~A. \& {Wilms}, J. 1999, \apj, 522, 476

\bibitem[{{Paredes-Fortuny} {et~al.}(2012){Paredes-Fortuny}, {Rib{\'o}},
  {Fors}, \& {N{\'u}{\~n}ez}}]{Paredes-Fortuny12}
{Paredes-Fortuny}, X., {Rib{\'o}}, M., {Fors}, O., \& {N{\'u}{\~n}ez}, J. 2012,
  in AIP Conf. Proc. 1505, High-Energy Gamma-ray Astronomy, ed. F.~A. {Aharonian}, W.~{Hofmann},
  \& F.~M. {Rieger} (Melville, NY: AIP), 390

\bibitem[{{Plotkin} {et~al.}(2013){Plotkin}, {Gallo}, \& {Jonker}}]{Plotkin13}
{Plotkin}, R.~M., {Gallo}, E., \& {Jonker}, P.~G. 2013, \apj, 773, 59

\bibitem[{{Remillard} \& {McClintock}(2006)}]{RemillardMcClintock06}
{Remillard}, R.~A. \& {McClintock}, J.~E. 2006, \araa, 44, 49

\bibitem[{{Straizys} \& {Kuriliene}(1981)}]{Straizys81}
{Straizys}, V. \& {Kuriliene}, G. 1981, \apss, 80, 353

\bibitem[{{Williams} {et~al.}(2010){Williams}, {Gies}, {Matson}, {Touhami},
  {Grundstrom}, {Huang}, \& {McSwain}}]{Williams10}
{Williams}, S.~J., {Gies}, D.~R., {Matson}, R.~A., {et~al.} 2010, \apjl, 723,
  L93

\bibitem[{{Zdziarski}(2012)}]{Zdziarski12b}
{Zdziarski}, A.~A. 2012, \mnras, 422, 1750

\bibitem[{{Zdziarski} {et~al.}(1998){Zdziarski}, {Poutanen}, {Mikolajewska},
  {Gierlinski}, {Ebisawa}, \& {Johnson}}]{Zdziarski98}
{Zdziarski}, A.~A., {Poutanen}, J., {Mikolajewska}, J., {et~al.} 1998, \mnras,
  301, 435

\end{thebibliography}

\end{document}